# OPPORTUNITIES FOR TECHNOSIGNATURE SCIENCE IN THE ASTRO2020 REPORT

By the NExSS Working Group on Technosignatures

Jacob Haqq-Misra (Blue Marble Space Institute of Science)
Sofia Sheikh (SETI Institute)
Manasvi Lingam (Florida Institute of Technology)
Ravi Kopparapu (NASA Goddard Space Flight Center)
Adam Frank (University of Rochester)
Jason Wright (Pennsylvania State University)
Eric Mamajek (Jet Propulsion Laboratory, California Institute of Technology)
Nick Siegler (Jet Propulsion Laboratory, California Institute of Technology)
Daniel Price (University of California: Berkeley)

The Astro2020 report outlines numerous recommendations that could significantly advance technosignature science. Technosignatures refer to any observable manifestations of extraterrestrial technology, and the search for technosignatures is part of the continuum of the astrobiological search for biosignatures (National Academies of Sciences 2019a,b). The search for technosignatures is directly relevant to the "World and Suns in Context" theme and "Pathways to Habitable Worlds" program in the Astro2020 report. The relevance of technosignatures was explicitly mentioned in "E1 Report of the Panel on Exoplanets, Astrobiology, and the Solar System," which stated that "life's global impacts on a planet's atmosphere, surface, and temporal behavior may therefore manifest as potentially detectable exoplanet biosignatures, or technosignatures" and that potential technosignatures, much like biosignatures, must be carefully analyzed to mitigate false positives. The connection of technosignatures to this high-level theme and program can be emphasized, as the report makes clear the purpose is to address the question "Are we alone?" This question is also presented in the Explore Science 2020-2024 plan[1] as a driver of NASA's mission.

This white paper summarizes the potential technosignature opportunities within the recommendations of the Astro2020 report, should they be implemented by funding agencies. The objective of this paper is to demonstrate the relevance of technosignature science to a wide range of missions and urge the scientific community to include the search for technosignatures as part of the stated science justifications for the large and medium programs that include the Infrared/Optical/Ultraviolet space telescope, Extremely Large Telescopes, probe-class far-infrared and X-ray missions, and various facilities in radio astronomy.

---

[1] https://science.nasa.gov/science-red/s3fs-public/atoms/files/2020-2024_Science.pdf

# Infrared/Optical/Ultraviolet (IR/O/UV) Space Telescope

The Astro2020 report recommends a flagship Infrared/Optical/Ultraviolet (IR/O/UV) space telescope to search for habitable planets (as well as to support general astrophysics). The main text of the report only refers to biosignatures, but the appendix on biosignatures (E-1) mentions technosignatures twice as a science objective of the IR/O/UV telescope. The Astro2020 report focused on biosignature capabilities, thus an Infrared/Optical/Ultraviolet (IR/O/UV) space telescope would also have some technosignatures capabilities.

Industrial pollution represents a class of atmospheric constituents on Earth that could conceivably be technosignatures if observed in the spectra of an exoplanet. One example is nitrogen dioxide ($NO_2$), which has large sources on Earth from combustion that are greater than non-anthropogenic sources. A study by Kopparapu et al. (2021) showed that the absorption features of $NO_2$ in the 0.2 −0.7 μm range could be detectable with the Large Ultraviolet Optical Infrared Surveyor (LUVOIR, Fischer et al. (2019)). Kopparapu et al. (2021) found that a 15 m LUVOIR-like telescope could detect Earth-like levels of $NO_2$ for a planet around a Sun-like star at 10 pc with ~400 hours of observation, while planets orbiting K-dwarf stars would require even less time due to the reduction in loss of $NO_2$ from photolysis in such systems. The detection of elevated $NO_2$ levels in the atmosphere of an exoplanet could be consistent with ongoing industrial processes on the surface, although any such observations would need to be evaluated against possible non-technological explanations before concluding that the $NO_2$ must be a technosignature. In this regard, such a search for technosignatures shares the same ambiguity or tentative nature as many or most searches for biosignatures. The Astro2020 decadal survey recommended a ~6 m aperture for the IR/O/UV space telescope. This recommendation was intended to provide high-contrast imaging capable of observing a "robust sample of ~25 atmospheric spectra of potentially habitable exoplanets," which could provide numerous opportunities to search for atmospheric technosignatures. While the study by Kopparapu et al. (2021) considered a 15 m telescope, the capabilities of the IR/O/UV telescope are similar, although observation times with the latter might be larger than proposed in Kopparapu et al. (2021) study due to a smaller aperture size.

The IR/O/UV telescope could also place constraints on the prevalence of optical beacons and other pulsed laser signals. Optical beacons could provide a cost-effective means of directed communication between exoplanetary systems, which could be encoded and transmitted through rapid nanosecond pulses. The motivation for optical technosignature searches is thus to place limits on the possibility of intentionally constructed optical or infrared laser beacons that were designed for the purpose of signaling or communication. Space missions such as the IR/O/UV telescope could provide detectability constraints on the prevalence of optical beacons in exoplanetary systems. For example, Vides (2019) calculated that the Gemini Planet Imager could detect a 24 kW laser from outside the habitable zone of τ Ceti while the Roman Space Telescope could detect a 7.3 W laser within τ Ceti's habitable zone. The latter result suggests relatively low-powered optical beacons could be detectable with the IR/O/UV telescope for most or all targets where the characterization of rocky planets within the HZ is also possible. Further study is needed to better understand the technosignature capabilities of the IR/O/UV telescope.

## Extremely Large Telescopes

The GMT and TMT are ongoing projects that have been developing for many years, and the Astro 2020 report recommended "significant U.S. investment in the Giant Magellan Telescope (GMT) and Thirty Meter Telescope (TMT) projects, ideally as components of a coordinated U.S. Extremely Large Telescope Program (ELT) program." These ground based facilities could be capable of characterizing the atmospheres of terrestrial planets discovered by missions like TESS and CHEOPS at optical and near-infrared wavelengths (López-Morales et al. 2019). Ground-based searches for atmospheric technosignatures in red dwarf systems may be possible before capable space-based missions launch. Possible spectral technosignatures such as atmospheric pollution and optical beacons were discussed above with relevance to the IR/O/UV telescope, and the prevalence of these classes of technosignatures could likewise be constrained with observations of exoplanetary systems by ELTs (López-Morales et al. 2019; Vides 2019; Kopparapu et al. 2021).

## Probe-class Far-infrared and X-ray Mission

The Astro2020 report recommended an "Astrophysics Probe Mission Program" at a cadence of one per decade, with the top two priorities being "a far-IR probe or an X-ray probe to complement the Athena mission." A far-IR mission could enable technosignature searches for megastructures and atmospheric pollution, while an X-ray mission could enable searches for novel anomalous signals.

*Far-infrared imaging or spectroscopy mission*

The significance of a far-infrared (far-IR) mission is apparent as early as Table S.5 (page S-8), where the importance of a Great Observatories Mission and Technology Maturation Program in the far-IR and X-ray wavelengths is underscored. Subsequently, Section 7.5.3.3 (page 7-20) emphasizes that a far-IR imaging or spectroscopy probe mission would "fill an important gap in world-wide capabilities". The Astro2020 decadal survey intimates that such a mission could potentially resemble the canceled ESA/JAXA far-IR mission entitled Space Infrared Telescope for Cosmology and Astrophysics (SPICA). The specifics of the far-IR probe were not discussed and would depend on the "outcome of the competitive selection." One example of a possible far-IR-focused probe-class mission is the Galaxy Evolution Probe (GEP), one of the probe mission concepts considered by Astro2020, which would have a spectrometer in the wavelength range of 24-193 μm.[2]

A far-IR mission clearly does not overlap with "traditional" search for artificial radio, optical, and near-IR narrowband signals. However, the aforementioned wavelength range is exciting for the so-called artifact SETI, of which the best-known example is Dyson Spheres, the energy-harvesting megastructures conceived by Olaf Stapledon and formalized by their eponym Freeman Dyson. Dyson spheres harvest a significant fraction of the radiation from the host star, but the laws of thermodynamics dictate that this energy is dissipated as waste heat. Dyson spheres would emit waste heat in the mid-infrared; for example, the emission from a Dyson sphere around a Sun-like star situated at a distance of 1 AU is predicted to peak at roughly 7 μm. However, a Dyson sphere would typically not have much far infrared emission, unlike dust. Thus, far IR capabilities offer a way to significantly reduce the problem of confounders such as protoplanetary disks.

Previous searches for Dyson spheres relied on data from the Infrared Astronomical Satellite (IRAS), with wavelengths ranging from 12 to 100 μm, and the Wide-field Infrared Survey Explorer (WISE) (see Wright (2020) and Lingam & Loeb (2021)). It is feasible in principle to replicate the construction of Dyson spheres on galactic scales, and one may therefore seek out signatures of waste heat from galaxies (extragalactic SETI, e.g., Wright et al. 2014; Griffith et al. 2015). When compared to previous surveys, a probe-class mission like the GEP would be three orders of magnitude more sensitive than IRAS and would likewise represent a substantial advance in sensitivity compared to WISE. It would thus be able to place more stringent constraints on the prevalence of Dyson spheres in our Galaxy, and also advance the prospects for extragalactic technosignatures.

Lastly, a number of interesting atmospheric technosignatures evince spectral features that are close to the wavelength range introduced previously. To offer one striking example, none of the abiotic or biological (but non-technological) pathways operating today can give rise to chlorofluorocarbons (CFCs), which have residence times of ~10 yr to ~$10^5$ yr. One of the strongest absorption features of CFC-11 ($CCl_3F$) is manifested at 11.7 μm, which is in proximity to the lower bound of 12 μm delineated earlier. Hence, a mid- to far-IR probe could conceivably be harnessed to search for atmospheric technosignatures such as CFCs, likely through transit spectroscopy as direct detection may be difficult for a small diameter mission.

Lin et al. (2014) predicted that an integration time of ~1 day might suffice to detect certain CFCs on a terrestrial planet in the habitable zone of a white dwarf by the James Webb Space Telescope (JWST), provided that the concentration of CFCs is about 10 times the present-day value. Likewise, Haqq-Misra et al. (2022) determined that the JWST could detect current or historic levels of CFCs in the atmosphere of the well-known temperate planet TRAPPIST-1e with ~ 100 hr of integration time to achieve a signal-to-noise ratio of ~3-5. We caution, however, that these analyses focused on the Mid Infrared Instrument (MIRI) and not far-IR spectral features.

---

[2] https://smd-prod.s3.amazonaws.com/science-red/s3fs-public/atoms/files/GEP_Study_Rpt.pdf

*X-ray Probe to Complement ESA's Athena Observatory*

In the same locations where the necessity and utility of a far-IR probe was foregrounded, the benefits accruing from an X-ray probe were also described. For instance, Table S.5 (page S-8) discusses the instantiation of the Great Observatories Mission and Technology Maturation Program at X-ray wavelengths. This notion is expanded in Section 7.5.3.4 (page 7-20), where it is argued that an X-ray probe would be vital and that it would complement ESA's Athena X-ray mission. The proposed X-ray probe is implicitly recommended to be equipped with "high spectral resolution (R ~ 7500), broad bandpass, and spatial (<1") resolution"; moreover, it may operate at X-ray energies of 0.1-8 keV if it resembles the Athena X-ray observatory.

At first glance, it would seem as though X-ray photons are not a promising "messenger" for artificial signals from ETIs since the latter are conventionally associated with radio (and optical) wavelengths. However, it has been proposed - based on considerations of maximizing the transmitted data rate - that narrowband X-ray emission centered at energies of 0.5-2 keV is optimal for communication (Hippke & Forgan 2017). This is because the photons can be focused towards Earth most tightly at higher wavelengths, and because this allows for the incorporation of timing into the information transmission. If this premise is correct, then some of the proposed probes with X-ray capabilities such as STROBE-X, HEX-P, TAP listed in Table J.2 may be well-suited for carrying out searches for these types of signals, although this topic warrants further investigation.

Aside from transmitting deliberate narrowband X-ray signals, there are other intriguing avenues whereby technologically advanced ETIs can broadcast their presence at X-ray wavelengths (Lingam & Loeb 2021). If a km-sized rock were to be hurled onto the surface of a neutron star, it may result in an intense X-ray pulse of ~$10^{29}$ W that might be detectable throughout the Milky Way. Another conceivable option is to take advantage of natural astrophysical X-ray sources such as low-mass X-ray binaries and apply suitable modulation techniques (e.g., by using lenses to boost brightness) that give rise to artificial patterns emblematic of ETI activity (Corbet 1997; Lacki 2020).

The chief benefit of the technosignatures outlined above is that they do not require the construction of specialized X-ray transmitters, and are anticipated to be less costly in energetic terms. Therefore, commensal searches for these types of phenomena may be readily carried out using the X-ray probe recommended by the Astro2020 decadal survey.

# Radio Astronomy

Technosignature science were mentioned a few times in the Radio, Millimeter, and Submillimeter (RMS) sections of Astro2020. In the main report text, SETI was recognized as one of the fundamental contributions of the Arecibo telescope. In the report from the RMS panel (Appendix M), technosignatures were mentioned twice: once as a way for time-domain searches with instruments like the DSA-2000 to contribute to the search for life on exoplanets, and once as a use-case for commensal backends on individual telescopes.

*Next Generation Very Large Array*

The RMS panel's priorities for the next decade align exceptionally well with the needs of technosignature science in the radio frequencies. The panel's top priority is the development and construction of the next generation Very Large Array (ngVLA), a recommendation that is echoed in the main report which states that "it is of essential importance to astronomy that [ngVLA] be built" [Chap. 7]. The ngVLA concept is a 244 antenna radio interferometer of unprecedented scale, supplying: an order-of-magnitude increase in sensitivity over its predecessors (the Jansky Very Large Array (JVLA) and the Very Long Baseline Array (VLBA)), sub-milliarcsecond angular resolution, high-fidelity (milliarcsecond-arcminute) imaging capabilities, and a bandwidth of 1.2-116 GHz.

Each of these state-of-the-art technical specifications makes ngVLA great for technosignature searches: the large number of small dishes leads to a fast sky survey speed, the sensitivity will allow for even unintentional "leakage" radio transmissions to be detected from nearby Earth-level transmitters, the angular resolution and imaging capabilities will allow for astrometric determinations of transmitter orbital motion within ~100 pc (adding another discriminant against radio frequency interference (RFI)), and the 10-100 GHz region is underexplored in the existing SETI literature, opening up a huge new swath of parameter space (Croft et al. 2018).

We envisage a ngVLA technosignature program in the style of Croft et al. (2018), where the main search mode is commensal with most primary science users of the array, but there is also a supplemental amount (10s-100s of hours) of primary-observer, TAC-allocated time for targeted technosignature searches. The commensal mode will provide stringent limits on radio transmissions from millions of targets, including stars, galaxies, and solar system objects, while the primary-observer time can be used for time-critical studies or follow-up of events of interest.

*Cosmic Microwave Background Stage 4 Experiment*

The other primary facility recommendation from the RMS panel was for the development of cosmic microwave background stage 4 experiment (CMB-S4), which will perform an ultra-deep survey of a few percent of the sky over 30-270 GHz with the goal of better understanding our cosmic web via the cosmic microwave background. CMB-S4 will facilitate ancillary science via the novel data products that it will produce in the 1cm--1mm wavelength range. It will create well-calibrated maps of ~50% of the sky with an every-other-day sampling rate, generate transient alerts for the community, and overall allow for "systematic time-domain studies in this part of the electromagnetic spectrum for the first time" [Chap. 7]. Commensal opportunities for these data products are many and varied (i.e., Eftekhari et al. 2021), and include technosignature searches, which have never been done in this corner of parameter space; CMB-S4 will primarily provide higher frequencies than any prior radio/microwave technosignature searches, but also provides a high repetition rate and thorough sky coverage.

*Mid-Scale Radio Instrumentation Projects*

Astro2020 highlights the importance of developing novel radio instrumentation via mid-scale projects, with a particular emphasis on radio transient cameras which can survey the static and time-variable sky. As described in Chapter 5, radio transient cameras "blend modern high-speed digitization and correlation of multiple resolution elements with optimized hardware to simultaneously monitor large swaths of sky." The RMS panel report also touts the development and deployment of new instrumentation, with accompanying data processing software, as a key investment.

SETI has always been, and continues to be, on the forefront of radio instrumentation development (i.e., Price et al. 2021; Welch et al. 2017). New instruments unlock new areas of parameter space, allow for commensal use of data, and provide new capabilities to existing facilities, all advantageous benefits for radio technosignature searches. In particular, proposed radio cameras such as the DSA-2000 or phased-array feeds on single dish telescopes like the GBT could be instrumental to technosignature searches via time-domain surveys to "potentially [search for] technosignatures as tracers of life on exoplanets (EAS-D)" (Appendix M), and are important in the quest for "all-sky all-the-time" monitoring of the radio sky for technosignatures (Elkers et al 2002).

*Pulsar timing, instrument development, and RFI Mitigation*

The RMS panel also promoted continuing support for 1) long-term pulsar timing, 2) development of new hardware and software, and 3) mitigation of RFI not tied to single facilities. Here we shall focus on the latter two recommendations, which align well with priorities for radio technosignature searches. The importance of the development of new hardware has already been discussed, but technosignatures especially benefit from the additional recognition of software as an area of radio observation that is ripe for development.

Technosignature searches are often framed as searches for a "needle in a haystack" of both parameter space and big data. In addition, the technosignature field is exploring machine learning methods, such as anomaly detection, in an effort to develop non-parametric search methods to mitigate anthropocentric assumptions (Berea et al. 2019). Software development in SETI has historically had a positive symbiotic relationship with other areas of astronomical software development (i.e., the relationship between technosignature search codes and pulsar dedispersion codes such as described by Taylor (1974)). Other areas of synergy include the use of computational methods from the pulsar and fast radio burst community for technosignatures application (e.g., Zackay and Ofek 2017; Gajjar et al. 2021).

As mentioned above, technosignature algorithms often rely on finding "anomalies" which are not consistent with known astrophysical emission mechanisms. This makes them fantastic at detecting and classifying RFI (which correspond to terrestrial technosignature signals). In addition, technosignature backends and data products have the finest frequency resolution of any microwave instrumentation (due to the prevalence of narrowband transmissions in human technology, which we extrapolate to technosignatures from extraterrestrial intelligence), providing an opportunity for high-quality RFI spectral density monitoring from the hardware side. Future technosignature searches will "provide a superb dynamic catalog of RFI, classified using machine learning algorithms, publicly accessible to other science users" (Croft et al. 2018), aligning perfectly with the RMS panel encouragement to "provide adequate funding to all current and future RMS facilities for the development of RFI protection and mitigation strategies, including specially designed hardware and sophisticated software tools to excise RFI" (Appendix M).

## Other Facilities

The Astro2020 recommendations establish priority for US participation in Cherenkov Telescope Array (C-Q2, L.4.3). This could provide further opportunities for the search for technosignatures, particularly the search for pulsed optical signals (Eichler & Beskin 2001; Abeysekara 2016).

## State of the Profession

Finally, the most vital investment that the decadal survey can recommend is an investment in the people who make up the profession, "without whom the ambitious facilities, instruments, and experiments, as well as the promised transformative discoveries, would lie unfulfilled." Technosignature searches, due to perennial underinvestment, need additional support more keenly than almost any other subfield in astronomy. Therefore, technosignatures will benefit hugely from the RMS panel's "support for traditional individual investigator grants", including those that support the training of graduate students, postdoctoral researchers, and "the next generation of instrument builders." In addition, technosignatures need the proposed "opportunities for small teams to pursue ambitious observing programs," and "facilities that offer opportunities for student training and/or substantial amounts of open time for diverse, risk-tolerant investigations" (Appendix M). It is worth noting that the astrobiological search for biosignatures and technosignatures has been shown to inspire diverse audiences and can serve as a catalyst for inspiring careers in STEM (see e.g., Impey 2021).

## Recommendations

The summary of technosignature capabilities in this paper is intended to highlight the opportunities should the recommended projects and programs recommended by the Astro 2020 Decadal Survey be realized. Although technosignatures were not explicitly mentioned in the main text of the report, they were discussed in Appendices E and M. The implementation of these recommendations would represent a significant advance in technosignature science capabilities. Technosignature observations can often be conducted commensally with other observations, and many technosignature searches can be conducted without changing the recommended mission architecture. The next generation of astrophysical facilities will therefore provide new and unprecedented opportunities to constrain the prevalence of technosignatures in exoplanetary systems.

This white paper recommends that all of the missions and facilities discussed above should consider including the search for technosignatures as part of the explicitly stated science case. Technosignature science will likely add no additional expense to such projects, and the science returns of finding technosignatures could be tremendous. This is not a call for technosignatures science to drive any of the Decadal recommended projects. Instead, this is a call to consider including technosignatures as ancillary science—by taking advantage of the capabilities of the Decadal-recommended projects—as an alternative means to "answer the fundamental question: 'Are we alone?'" in the Priority Area "Pathways to Habitable Worlds."

### Acknowledgements

The results reported herein benefitted from collaborations and/or information exchange within NASA's Nexus for Exoplanet System Science (NExSS) research coordination network sponsored by NASA's Science Mission Directorate. Part of this research was carried out at the Jet Propulsion Laboratory, California Institute of Technology, under a contract with the National Aeronautics and Space Administration (80NM0018D0004). This research was also supported in part by a grant from the NASA Exobiology program (80NSSC20K0622).